\documentclass[a4paper,11pt]{article}
\usepackage{pos_1}

\title{A proposal for experiment with high-intensity tritium neutrino source in Sarov: The search for coherent elastic neutrino-atom scattering and neutrino magnetic moment}
\ShortTitle{The search for neutrino-atom scattering and neutrino magnetic moment in Sarov}

\author[a]{Matteo Cadeddu}
\author[b]{Georgy Donchenko}
\author[a]{Francesca Dordei}
\author[c]{Carlo Giunti}
\author[b]{Konstantin Kouzakov}
\author[d]{Bayarto Lubsandorzhiev}
\author*[b]{Alexander Studenikin}
\author[e]{Vladimir Trofimov}
\author[b]{Maxim Vyalkov}
\author[g]{Arkady Yukhimchuk}
\affiliation[a]{Istituto Nazionale di Fisica Nucleare, Sezione di Cagliari,\\
  Complesso Universitario di Monserrato - S.P. per Sestu Km 0.700, 09042 Monserrato (Cagliari), Italy}
\affiliation[b]{Faculty of Physics and MSU Branch in Sarov (NCPM), Lomonosov Moscow State University,  Moscow 119991, Russia}
\affiliation[c]{Istituto Nazionale di Fisica Nucleare, Sezione di Torino,
Via P. Giuria 1, I–10125 Torino, Italy}
\affiliation[d]{Institute of Nuclear Research of the Russian Academy of Sciences,
Moscow 117312, Russia}
\affiliation[e]{Joint Institute for Nuclear Research,
Dubna 141980, Moscow Region, Russia}
\affiliation[g]{National Centre for Physics and Mathematics,
Sarov, Nizhni Novgorod Region, Russia}

\emailAdd{matteo.cadeddu@gmail.com}
\emailAdd{dongosha@gmail.com}
\emailAdd{francesca.dordei@cern.ch}
\emailAdd{carlo.giunti@to.infn.it}
\emailAdd{kouzakov@gmail.com}
\emailAdd{lubsand@rambler.ru}
\emailAdd{studenik@srd.sinp.msu.ru}
\emailAdd{trof@jinr.ru}
\emailAdd{arkad@triton.vniief.ru}
\emailAdd{vialkov.mm17@physics.msu.ru}

\abstract{A description of the current state of the project for the study of coherent elastic neutrino-atom scattering using a tritium source and liquid helium detector is given. The project was proposed in our paper in 2019 and its main goal is to obtain a new record limit on the neutrino magnetic moment at a level below $10^{-12}\mu_B$ using a tritium antineutrino source with an intensity of 10 MCi or even 40 MCi.}

\FullConference{%
  41st International Conference on High Energy physics - ICHEP2022\\
  6-13 July, 2022\\
  Bologna, Italy
}

\begin{document}
\maketitle

\section*{Introduction}
The studies of neutrino electromagnetic properties can open a window to new physics \cite{Giunti:2014ixa}. This is because the neutrino magnetic moments, the most well understood and studied among neutrino electromagnetic characteristics, are zero in the Standard Model with massless neutrinos.  In a minimal extension of the Standard Model  the diagonal magnetic  moment of a massive Dirac neutrino $\nu_i$ is given \cite{Fujikawa:1980yx} by
$\mu^{D}_{ii}
  = \frac{3e G_F m_{i}}{8\sqrt {2} \pi ^2}\approx 3.2\times 10^{-19}
  \Big(\frac{m_i}{1 \ \mathrm{eV} }\Big) \mu_{B}$,
where $\mu_B$ is the Bohr magneton. The strongest upper bounds for the neutrino magnetic moments in laboratory experiments are obtained with reactor neutrinos:
$\mu_{\nu} \leq 2.9 \times 10^{-11} \mu_{B}$ (GEMMA Collaboration \cite{GEMMA:2012}), and solar neutrinos:
${\mu}_{\nu}\leq 2.8 \times
10^{-11} \mu _B$ (Borexino Collaboration \cite{Borexino:2017fbd}). An order of magnitude stronger constraints are provided by studies of astrophysical neutrinos (see for example  \cite{Giunti:2014ixa}). Thus, it is an important goal of the present fundamental physics research to fill the gap between the present experimental limits and the theoretical predictions for the neutrino magnetic moments.

In our recent paper \cite{Cadeddu:2019qmv}  we have proposed an experimental setup to observe coherent elastic neutrino-atom scattering ($CE \nu AS$) using electron antineutrinos from tritium decay and a supefluid $^4$He target. In this scattering process with the whole atom, that has not been observed so far, the electrons tend to screen the weak charge of the nucleus as seen by the electron antineutrino probe.

 		\vspace{-5mm}
	{\center{\includegraphics[width=0.8\linewidth]{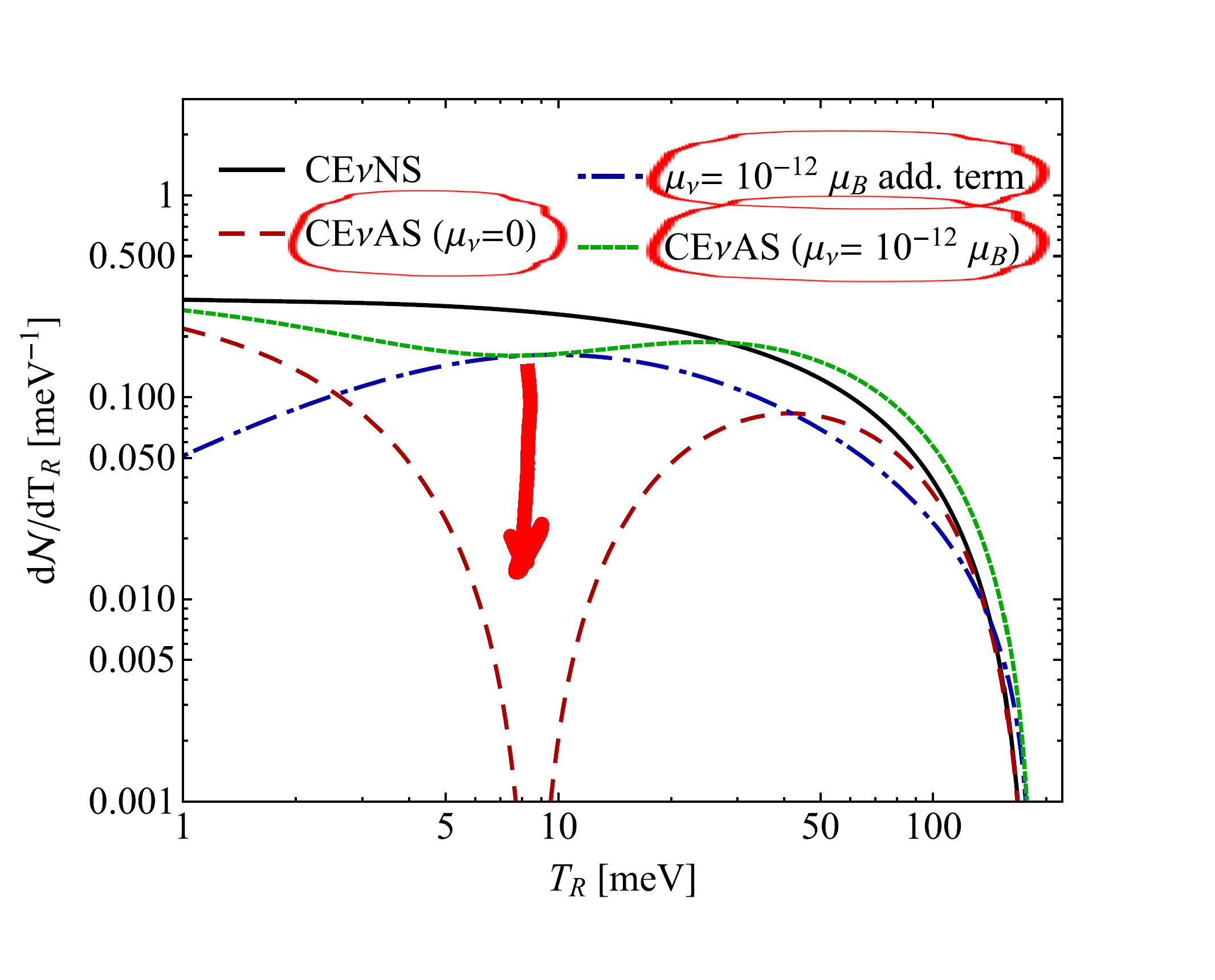}}
	\vspace{-10mm}
	\captionof{figure}{Atomic recoil spectrum.}
}
The differential number of neutrino-induced events as a function of the atomic recoil energy $T_R$ in a logarithmic scale is shown in Fig. 1.  The coherent elastic neutrino nucleus scattering ($CE \nu NS$) differential number is shown by the black solid line while the $CE \nu AS$ one is shown by the dashed red line. The dashed-dotted blue line represents the additional term appearing in the $CE \nu AS$ differential number of events
assuming a neutrino magnetic moment of $\mu_{\nu} = 10^{-12} \mu_B$ while the dotted green line represents the total differential number of
$CE \nu AS$ for the same value of $\mu_{\nu}$. We find that it is possible to set an upper limit 
that is about two orders of magnitude smaller than
the current experimental limits from GEMMA and Borexino. A corresponding experiment involving an intense tritium neutrino source is currently being prepared in the framework of the research program of the National Center for Physics and Mathematics in Sarov (Russia).
\section{Tritium source of neutrinos}
The advantages of a tritium source (TS) as compared to solar, nuclear, accelerator and other $\beta$-active sources may be summarized as follows~\cite{2}:
	1) more intensive (anti)neutrino fluxes as compared with the reactor and accelerator sources and
	strongly suppressed correlated background; 	2) small sizes that make it possible to locate an experiment in a moderate-size low-background underground laboratories and flux modulation for the non-correlated background deduction;
3) the (anti)neutrino spectrum defined with high accuracy; 4) a quite low maximum of the neutrino energy spectrum ($E_0 = 18.6$ keV), thus the bremsstrahlung radiation does not penetrate beyond the limits of the source and there is no necessity of a passive protection between the source and detector.
%

 To measure the magnetic moment of neutrinos, it was proposed in~\cite{2,3} to use a tritium source with an activity of 40 MCi (4 kg of tritium) (see Fig.~\ref{ris:experimcoded}).

		{\center{\includegraphics[width=0.5\linewidth]{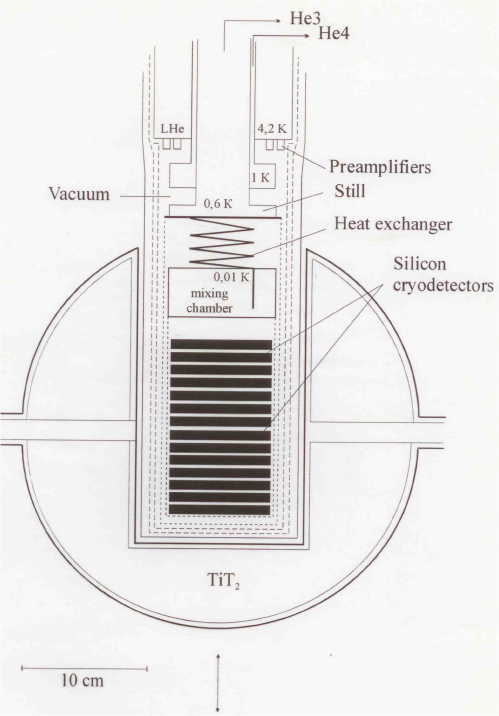}}
		\captionof{figure}{Experimental setup: an assembly of semiconductor detectors is located inside a spherical tritium source ($TiT_2$).}
		\label{ris:experimcoded}}
	



The tritium in the TS is in a chemically bound state with Ti, which provides an adsorption capacity of tritium of 1700 ${\text{cm}^{3}}/{\text{cm}^{3}}$, a low equilibrium pressure at room temperatures of $\sim 10^{-9}$ mbar, a high decomposition temperature of 550-620~$^{o}$C hydride and absence of radioactive impurities of the uranium series.

	{\center{\includegraphics[width=0.5\linewidth]{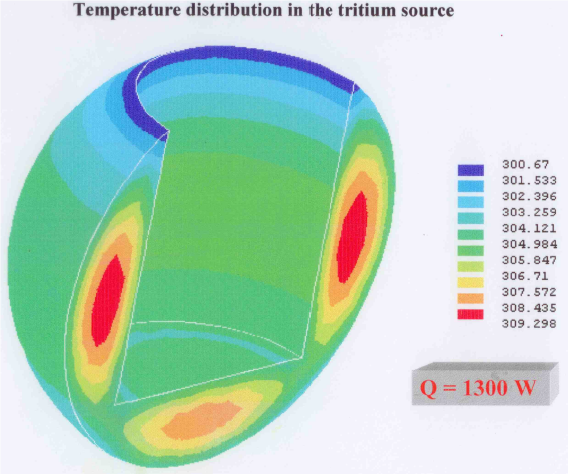}}
	\vspace{0mm}
	\captionof{figure}{The temperature distribution in the TS.}
	\label{fig02}}

In the works~\cite{2,3,4}, the possibility of a fundamental implementation of such an experimental scheme was shown. However, the thermal calculation of the TS, under conditions of natural storage, showed its significant heating. In the center of the source, the temperature exceeded 309~${}^{o}$C (see Fig.~\ref{fig02}), which caused certain difficulties during transportation, storage and operation of such a source.

In~\cite{5}, the TS design was optimized. The following factors were taken into account: the available amount of tritium; obtaining the maximum number of neutrino interaction events with the detector material; feasibility of manufacturing; safety at all stages of the TS life cycle; ease of installation and disassembly.


Taking into account the above factors, Fig.~\ref{ris:experimoriginal} shows the evolution of views on the design of TS.
The design scheme of the source is a cylindrical ring assembly of 80 tubular elements (see Fig.~\ref{TE}) filled with titanium tritide. Detectors are placed inside the ring assembly. Advantages of such a design: free access to a specific tubular element during its installation/disassembly; easily removal of the heat from the decay of tritium; the transportation of each of a single fuel element is carried out within the framework of existing international and Russian standards.
\vspace{-5mm}
\begin{center}
	\begin{minipage}[h]{0.45\linewidth}
		\center{\includegraphics[width=1.2\linewidth]{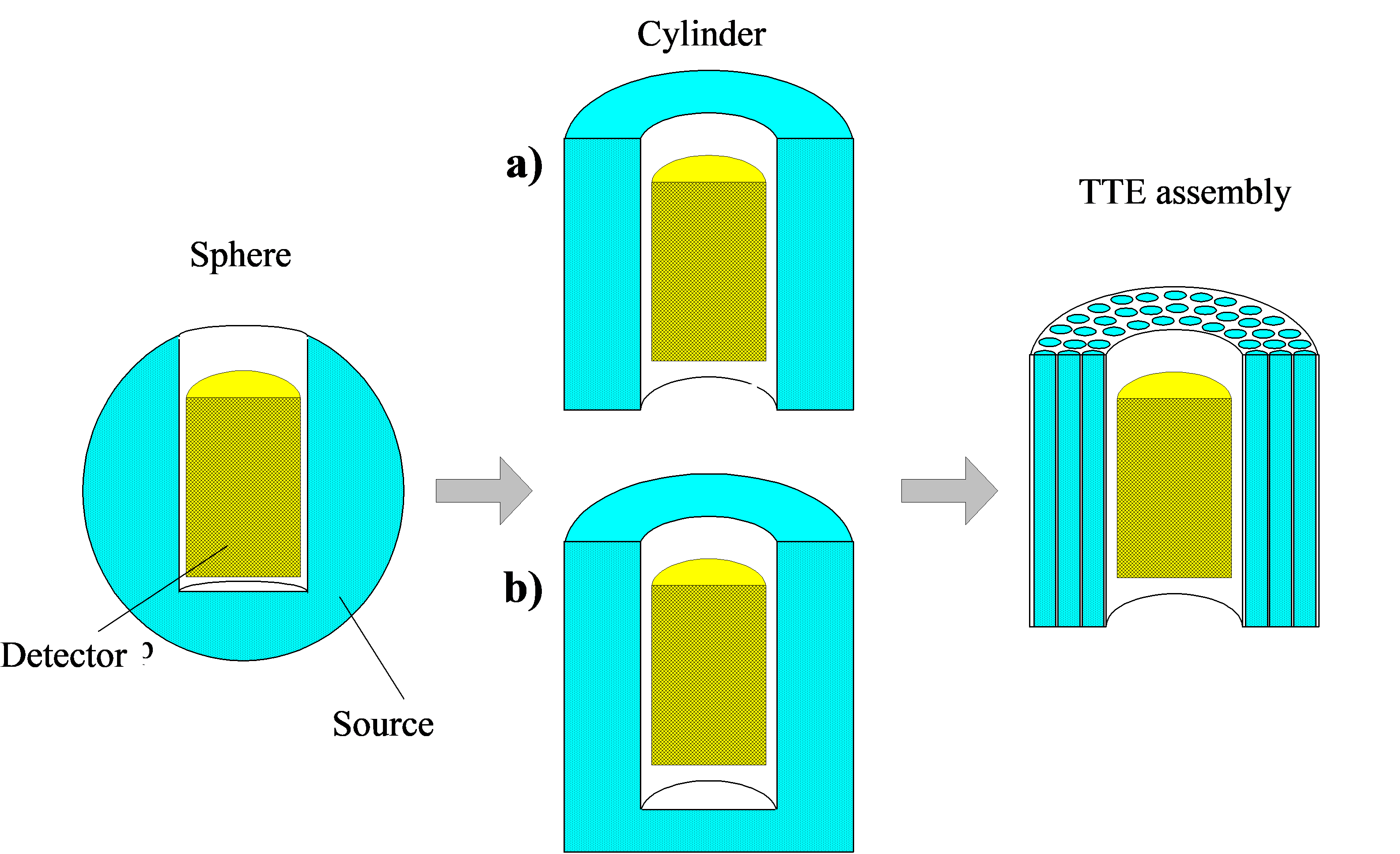}}
		\captionof{figure}{The evolution of designs of the TS.}
		\label{ris:experimoriginal} 
	\end{minipage}
	\hfill
	\begin{minipage}[h]{0.4\linewidth}
		\center{\includegraphics[width=0.18\linewidth]{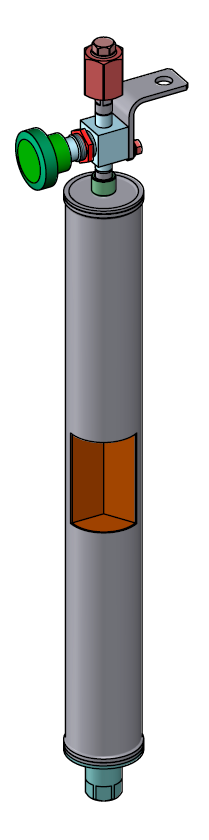}}
		\captionof{figure}{The tubular element.}
		\label{TE}
	\end{minipage}
\end{center}
	
	%
	%
%
	%
	%
	%
	\section{Experimental scheme}

We consider a detector setup such that the tritium source
is surrounded with a cylindrical superfluid-helium tank, as
depicted in Fig.~\ref{detector_layout}. This configuration allows us to maximize
the geometrical acceptance, while allowing us to have
a top flat surface where helium atoms could evaporate after
a recoil.

	{\center{\includegraphics[width=0.3\linewidth]{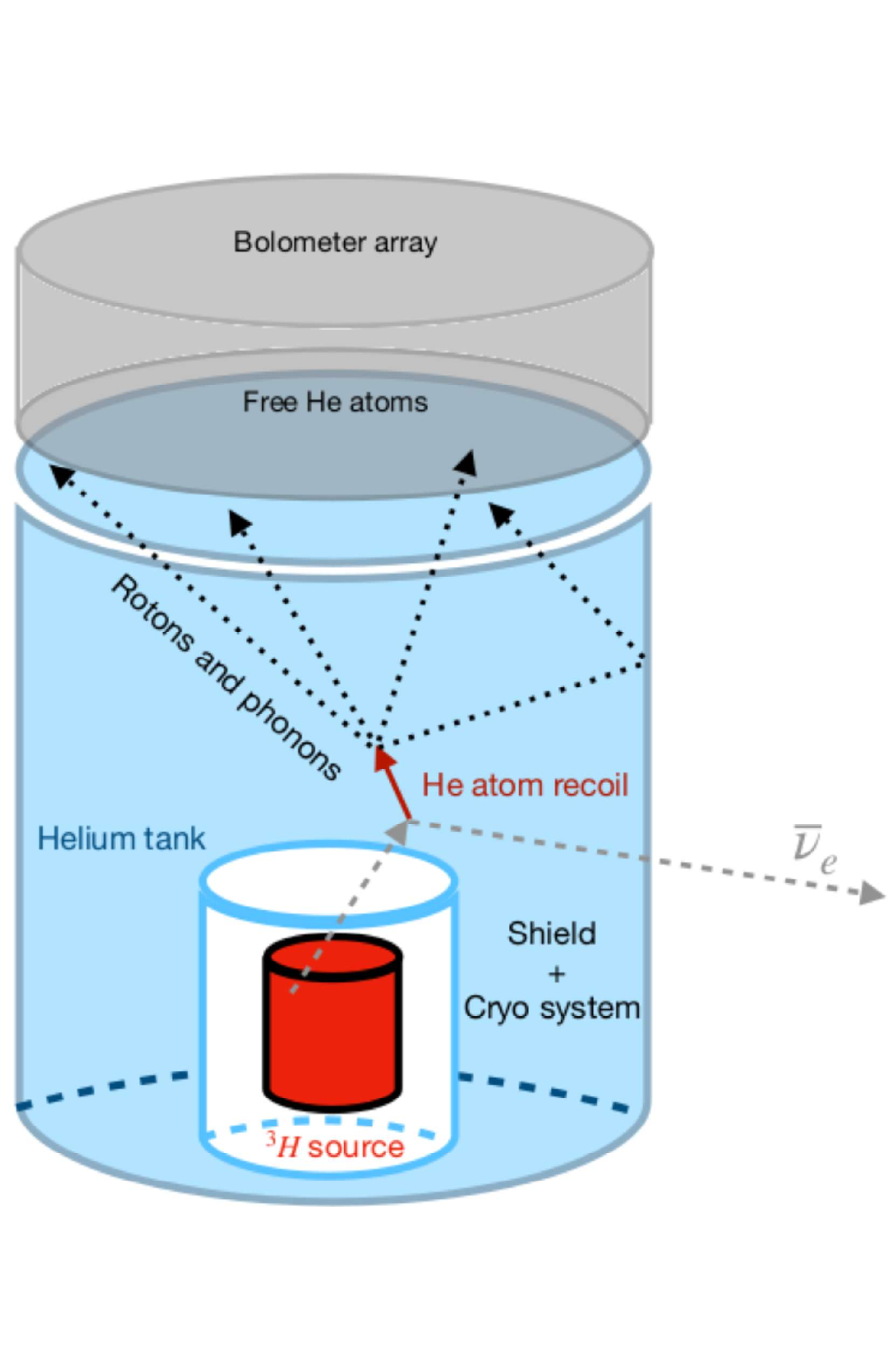}}
	\captionof{figure}{Schematic representation of the detector proposed to
observe CE$\nu$AS processes. The recoil of a helium atom after the
scattering with an electron antineutrino coming from the tritium
source produces phonons and rotons which, upon
arrival at the top surface, cause helium atoms to be released by
quantum evaporation. An array of bolometers on the top
surface detects the number of helium
atoms evaporated.}\label{detector_layout}}

\section{Expected number of events}
%
The source and tank dimensions depend on the tritium source activity. 
We consider the tank's volume of about 1 m$^3$ and two values of the tritium source activity: 10 MCi (minimally expected) and 40 MCi (maximally expected).
		%
	
The expected number of the CE$\nu$AS events for 5 yr of data taking are $N^{\rm CE\nu AS}=58.9$ and $195.2$ for the tritium source activity of 10 and 40 MCi, respectively.	These numbers allow us to reach the sensitivity to CE$\nu$AS of at least 5$\sigma$. A successful implementation of such scheme will also allow us to test the neutrino magnetic moment at the level of $3.8 \cdot 10^{-13} \mu_B$ and $2.3 \cdot 10^{-13} \mu_B$ at $90 \%  \ C.L.$ for 10 MCi and 40 MCi tritium source, respectively.

%
	%
	
	\section{Prospects}
	
	

The proposed experiment with a superfluid $^4$He target and a tritium neutrino source will allow us to study for the first time the CE$\nu$AS process and to test the neutrino magnetic moment and other neutrino electromagnetic properties (millicharge and charge radius) at an unprecedented high level of sensitivity. We plan to start the measurements in 2026 in a low-background neutrino lab that will be established in Sarov for this purpose. We expect to use the tritium source with the activity of at least 10 MCi (1 kg of tritium) and the amount of tritium can be increased to reach 40 MCi (4 kg of tritium).

We also plan to develop and use a Si detector with a low-energy threshold for testing the neutrino magnetic moment and other neutrino electromagnetic properties by measuring elastic neutrino-electron scattering. The 20-kg Si detector is expected to operate at temperatures of few tens of mK and to have the energy threshold as low as $\sim10$ eV or even $\sim1$ eV due to the Neganov-Trofimov-Luke effect~\cite{6}.
With such a detector and the discussed tritium source, we expect to achieve a sensitivity to the neutrino magnetic moment at a level below $10^{-12}\mu_B$.

The work is supported by the Russian Science Foundation under grant No.22-22-00384.

\end{document}